\documentclass[twocolumn]{aastex63}
\usepackage{graphicx}
\usepackage{amssymb}
\usepackage{amsmath}
\usepackage{verbatim}
\usepackage{natbib} 
\begin{document}

\title{Ly$\alpha$ at Cosmic Dawn with a Simulated \textit{Roman} Grism
Deep Field}

\author[0000-0002-0784-1852]{Isak G. B. Wold}
\affil{Astrophysics Science Division, Goddard Space Flight Center, Greenbelt, MD 20771, USA}
\affil{Department of Physics, The Catholic University of America, Washington, DC 20064, USA }
\affil{Center for Research and Exploration in Space Science and Technology, NASA/GSFC, Greenbelt, MD 20771}

\author[0000-0002-9226-5350]{Sangeeta Malhotra} 
\affiliation{Astrophysics Science Division, Goddard Space Flight Center, Greenbelt, MD 20771, USA}

\author[0000-0002-1501-454X]{James E. Rhoads} 
\affiliation{Astrophysics Science Division, Goddard Space Flight Center, Greenbelt, MD 20771, USA}

\author[0000-0001-8514-7105]{Vithal Tilvi} 
\affiliation{School of Earth and Space Exploration, Arizona State University, Tempe, AZ 85287, USA}

\author[0000-0003-4295-3793]{Austen Gabrielpillai} 
\affiliation{Astrophysics Science Division, Goddard Space Flight Center, Greenbelt, MD 20771, USA}
\affil{Department of Physics, The Catholic University of America, Washington, DC 20064, USA }
\affil{Center for Research and Exploration in Space Science and Technology, NASA/GSFC, Greenbelt, MD 20771}

\begin{abstract}
The slitless grism on the \textit{Nancy Grace Roman
Space Telescope} will enable deep near-infrared
spectroscopy over a wide field of view. We demonstrate \textit{Roman}'s capability to detect Ly$\alpha$ galaxies at $z>7$ using a multi-position-angle (PA) observational strategy. We simulate
\textit{Roman} grism data using a realistic foreground scene from the COSMOS field. We also input fake Ly$\alpha$ galaxies spanning
redshift z=7.5-10.5 and a line-flux range of interest. We show how a novel
data cube search technique -- CUBGRISM -- originally developed for
\textit{GALEX} 
can be applied to \textit{Roman} grism data
to produce a Ly$\alpha$ flux-limited sample without the need for
continuum detections. We investigate
the impact of altering the number of independent PAs and
exposure time. A deep \textit{Roman} grism survey with 25 PAs and a total exposure time of $70$hrs can achieve Ly$\alpha$ line depths comparable to the deepest $z=7$ narrow-band surveys  ($L_{{\rm{Ly}}\alpha}\gtrsim10^{43}$erg s$^{-1}$).  Assuming a null result, where the opacity of the intergalactic medium (IGM) remains unchanged from $z\sim7$, this level of sensitivity will detect $\sim400$  deg$^{-2}$ Ly$\alpha$ emitters from $z=7.25-8.75$. A decline from this expected number density is the signature of an increasing neutral hydrogen fraction and the onset of reionization.  Our simulations indicate that a deep \textit{Roman} grism survey has the ability to measure the timing and magnitude of this decline, allowing us to infer the ionization state of the IGM and helping us to distinguish between models of reionization.

\vspace{1cm}
\end{abstract}

\section{Introduction}

Ly$\alpha$ emission is a bright spectral feature in star-forming
galaxies that many observational surveys have targeted to efficiently
obtain large samples of Ly$\alpha$ emitting galaxies over a wide
redshift range $z=2-7$ \citep[e.g.,][]{cowie98,rhoads00,malhotra02,ouchi03,gawiser07,gronwall07,blanc11,finkelstein13,konno14,matthee15,tilvi20}. With the arrival of JWST, Ly$\alpha$ continues to play an essential role in the study of high-redshift galaxies and the ionization state of the intergalactic medium (IGM) \citep{endsley22,tang23,bunker23,saxena23}.    The most exciting JWST Ly$\alpha$-observation to date is the discovery of an emitter at z=10.6 \citep{bunker23}. This object shows that Ly$\alpha$ can find a way to escape only $430$ Myr after the Big Bang, where even extremely early reionization models \citep[e.g.,][]{finkelstein19} predict a neutral hydrogen fraction of  $\sim80\%$. However, the current Ly$\alpha$ JWST observations are limited in volume and are pre-selected based on broadband criteria.  We need a deep, blind search for Ly$\alpha$ over a large volume to realize the full potential of Ly$\alpha$-based reionization constraints.

We know that reionization should fall within the $6<z<9$ redshift range from the saturation of Ly$\alpha$ absorbers in $z\sim6$ quasar spectra \citep{fan06} and by polarization measurements of the cosmic microwave background \citep{planck18}. To go beyond this rough picture, many groups have used the redshift evolution of the number density
of Ly$\alpha$ emitters (LAEs) to constrain the timing of reionization.
Some groups have found a large drop in their density at $z\sim7$
and have attributed this decline to the increasing opacity of the
IGM and the onset of the epoch of reionization
\citep[e.g.,][]{stark10,konno14,hoag19}, while other groups have
found a more modest evolution over this same redshift range consistent
with a largely ionized universe (e.g., \citealt[][]{tilvi10,itoh18,zheng17,hu19,wold22}). Some of the variation in the results may be due to spatially inhomogeneous reionization since cosmological reionization is thought to have proceeded through the growth and eventual overlap of ionized bubbles, driven by ultraviolet photon production in the earliest galaxies. This produces a non-uniform reionization history that necessitates observational studies with large survey volumes to mitigate cosmic variance.

The current variation and uncertainty in the observational constraints allow for a diverse set of reionization histories. These reionization scenarios range from a rapid neutral-to-ionized transition dominated by relatively massive galaxies \citep{naidu20} to a gradual transition dominated by more numerous low-mass galaxies \citep{finkelstein19}. The redshift where these reionization models differ the most is $z\sim8$. For example, at this redshift Naidu et al.\ predict an almost fully neutral universe, while the Finkelstein et al.\ model predicts a much lower neutral fraction of $35\%$.  Moving to lower (higher) redshifts the models converge to a fully ionized (neutral) IGM, making $z\sim8$ an ideal redshift to distinguish between models and tell us about the sources of ionization radiation - whether large galaxies or small.

However, it is difficult to obtain $z\sim8$ Ly$\alpha$-based observational constraints from ground facilities largely due to the
increasing night sky background and the need for expensive spectroscopic
followup of Ly$\alpha$ candidates to eliminate contamination. To effectively study reionization at $z\sim8$, we need to acquire greater
statistical leverage by obtaining
spectroscopically complete samples of LAEs over larger volumes without the interference of the Earth's atmosphere. 

In this paper, we investigate \textit{Roman}'s potential to obtain
a LAE sample at $z\gtrsim7$ with a multi-position-angle grism deep
field. \textit{Roman}'s Wide Field Instrument (WFI) has visible to
near-infrared imaging capability and a slitless grism capability spanning
a wavelength range of $1.00-1.93$ $\mu$m with an $\sim11$ \AA\
per pixel dispersion. The WFI has a field of view of $0.281$ square
degrees or over 100 times larger than similar instruments on \textit{HST}
and \textit{JWST}. This large field-of-view (FOV) is tiled by 18 detectors
in 3 rows and 6 columns with a $0.11''$ pixel scale.  It is the combination of \textit{Roman}'s wide field of view and grism wavelength range that will clear the way for $z>7$ LAE searches with unprecedented volumes  -- a volume of $\sim3\times10^{6}$ Mpc$^{3}$ from $z=7.25$ to $8.75$ in a single pointing.

Slitless grism capabilities are well suited for space-based telescopes where the sky background is low, and where slitless modes are often the simplest way to add spectroscopy to a mission.
Furthermore, slitless observations are extremely efficient at spectroscopic surveys because all objects within the field-of-view have their light dispersed across the detector.  This is particularly a desirable feature for finding strong emission line sources that are easily missed in slit-mask observations which often select their targets based on continuum brightness \citep{malhotra05,tilvi16,larson18}.  

These strengths also present challenges.  In particular, neighboring sources aligned in the dispersion direction can have overlapping spectra. This means that a high-redshift slitless grism survey will need to contend with contamination caused by the foreground scene of stars and galaxies, and a simple exposure time calculation will fail to capture the true survey performance given this source confusion.  For these reasons, we simulate a realistic extra-galactic foreground scene that captures the key characteristics of a grism deep field and then assess our ability to conduct a blind search for Ly$\alpha$ galaxies at cosmic dawn. We show how a novel data cube search technique -- CUBGRISM -- originally developed for
\textit{GALEX} grism data can be applied to \textit{Roman} data allowing us to isolate $z>7$ Ly$\alpha$ galaxies.

\section{Grism Simulations}

We use \textit{HST} imaging and best-fit SEDs in the COSMOS field
to define our simulated foreground stars and galaxies. We also construct
fake high-redshift ($7.5<z<10.5$) Ly$\alpha$ emitters with a variety
of intrinsic characteristics and then measure our ability to detect
these emitters with different observational strategies. In the following
sections, we describe the construction of our grism simulations in
detail.

\subsection{Foreground Scene}

Our simulated foreground objects are populated based on \textit{HST}
observations of the COSMOS field. Specifically, we use the version-4.1
3D-\textsl{HST} COSMOS F160W mosaic with the derived photometric and
EAZY output files that are publicly available via the 3D-\textit{HST}
collaboration \citep{skelton14,brammer12,grogin11,koekemoer11}. The \textit{HST} COSMOS field has a $\sim9\times22$ arcmin$^{2}$ field of view observed with 44 filter bandpasses including 5 \textit{HST} filters: F606W, F814W, F125W, F140W, and F160W with $5\sigma$ depths of 26.7, 26.5, 26.1, 25.5, and 25.8, respectively \citep{skelton14}.

\subsubsection{Morphology\label{morpho}}

We define the morphology of simulated objects with image postage stamps.
We use the 3D-\textsl{HST} F160W mosaic to make these images for all
of our simulated foreground sources. For each object, we extract
a sub-image from the science F160W image centered on the object's
coordinates, where all pixels not assigned to the object via the segmentation
map are zeroed out. To prepare these sub-images for input, we resample
the F160W mosaic's drizzled pixel scale of $0.06''$ to match \textit{Roman}'s
pixel scale of $0.11''$. Note, the native pixel scale of WFC3-IR is about $0.13''$/pixel, so we're producing a final model image with a pixel scale very near the actual input data. The F160W mosaic has a median $5\sigma$
depth of 25.8 mag and an average FWHM of $0.19''$ \citep{skelton14}.
Our foreground simulations include all F160W objects with magnitudes
brighter than 25.5 mag. This corresponds to the expected $1\sigma$ per pixel sensitivity at $1.5\mu$m for our deepest simulated
Roman grism survey with a total $t_{{\rm {exp}}}=69.4$ hrs\footnote{based on the Roman Reference Information: \url{https://roman.gsfc.nasa.gov/science/Roman_Reference_Information.html}}.

\subsubsection{Spectra}

\begin{figure}
\begin{centering}
\includegraphics[bb=0.905bp 0.961039bp 724bp 518bp,clip,height=6cm]{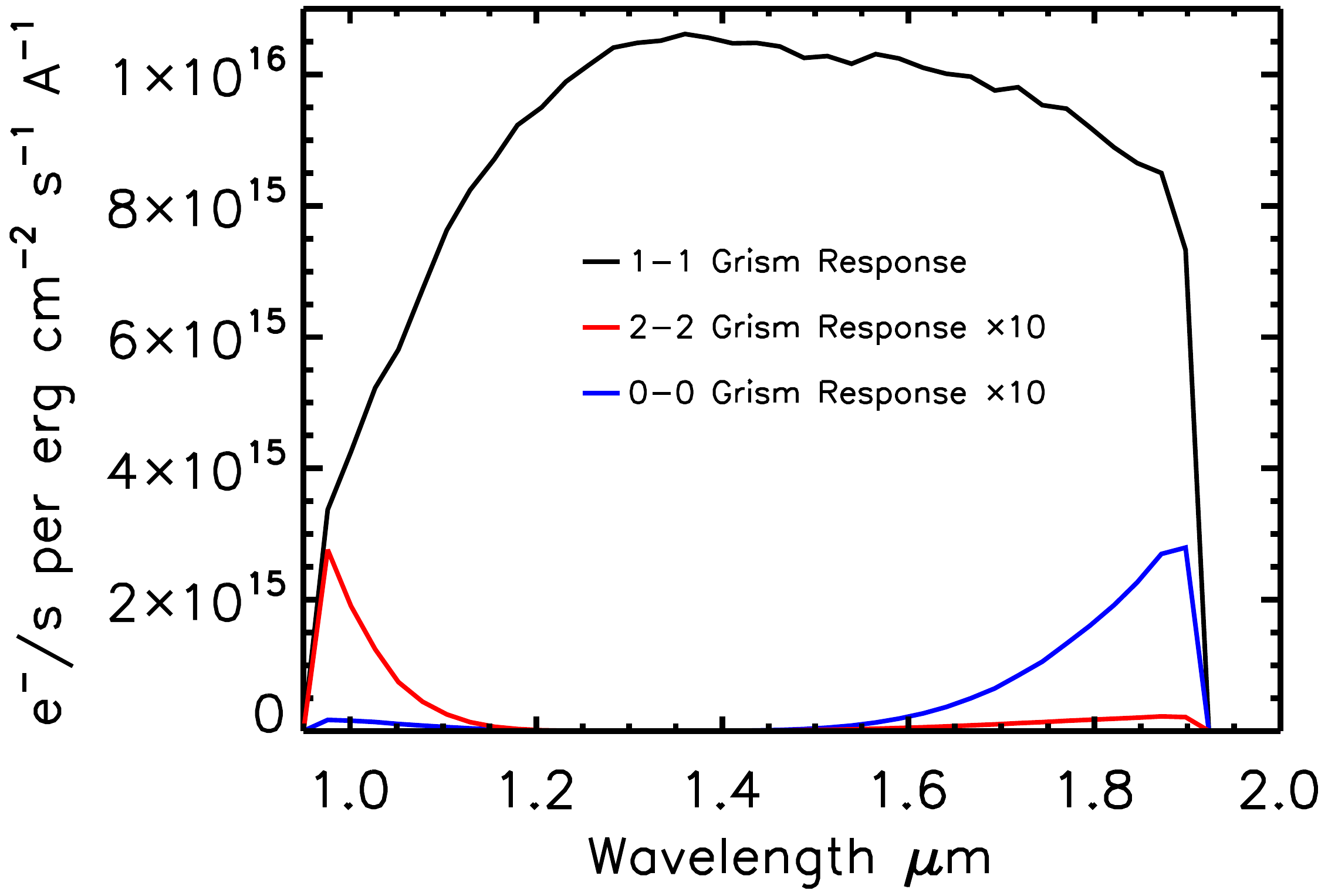}\caption{Roman WFI grism response functions for the in-focus spectral orders.
For display purposes, we increased the 2-2 and 0-0 grism response
functions by a factor of ten. All three in-focus spectral orders are included in our grism simulations.}\label{resp}
\par\end{centering}
\end{figure}

\begin{figure*}
\centering{}\includegraphics[bb=190.9365bp 0bp 959bp 185bp,clip,width=18cm]{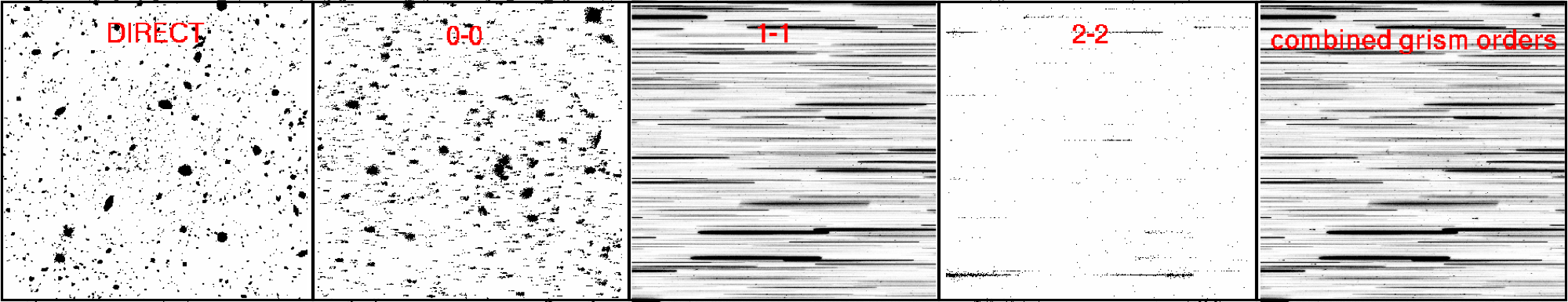}\caption{Construction of the noiseless simulated grism images. All the in-focus spectral orders: 0-0, 1-1, and 2-2 are simulated independently
and then co-added to form our combined grism image. Finally, we add
in the Poisson noise appropriate to our desired exposure time (see
Figure \ref{grism_fin67}). For the four panels above, we show a $3.75\times3.75'$ field of view.}
\label{construct67}
\end{figure*}

We use the best-fit EAZY SEDs from \citet{skelton14} to define the
spectra of all the simulated foreground objects. These spectral templates
were fit to 44 bandpasses with 13 spanning our simulated $1$ to $2$
$\mu$m wavelength range. Skelton et al.\ find excellent agreement
with their photometric redshifts and the available spectroscopic redshifts,
with a scatter of $\sigma_{{\rm {NMAD}}}=0.007$ and an outlier percentage
of $1\%$.  This level of agreement suggests that their templates SEDs, which we use directly, are able to accurately represent the spectral data.

For our \textit{Roman} grism simulations, we are most concerned
with reproducing a realistic foreground scene for the $1$ to $2$
$\mu$m wavelength range, and we normalize the EAZY SEDs to the observed
F160W photometry to help ensure that our simulated grism image has
a realistic distribution of foreground flux.

\subsection{Simulated Ly$\alpha$ emitters \label{lya}}

Simulated Ly$\alpha$ emitters are populated with a range of redshifts:
$z=\{7.5,8.5,9.5,10.5\}$, lines fluxes: $f_{{\rm {Ly}\alpha}}=\{0.45,0.6,1.0,1.2,4.0,7.0\times10^{-17}$
erg s$^{-1}$cm$^{-2}\}$, continuum magnitudes: $m_{{\rm {F160W}}}=\{25.5,26.5,27.5,28.5\}$,
and morphologies: half-light radii $=0.17\pm0.02''$. To achieve realistic Ly$\alpha$ morphologies, each object's semi-major and semi-minor axis lengths are uniformly selected over an interval spanning from 2 to 5 \textit{Roman} pixels.
This randomly selected size is then paired with an F160W cataloged source
with a matching spatial extent. This observed F160W source is then used
to generate the needed image postage stamp as discussed in Section
\ref{morpho}. The end result is distribution with $\left\langle r_{{\rm 1/2}}\right\rangle \pm1\sigma=0.17\pm0.02''$
which is consistent with the expected size of $z\gtrsim7$ LAEs \citep[][]{malhotra12, allen17, shibuya19,kim21,rhoads23}.

For the Ly$\alpha$ spectral shape, the continuum has a spectral slope
of $-2$ such that $f_{\lambda}\varpropto\lambda^{-2}$ and is attenuated
by the intervening IGM. The Ly$\alpha$ line is a Gaussian with rest-frame
FWHM$=1$\AA\ or 250 km s$^{-1}$ which is consistent with observational constraints \citep[][]{hu10}. We insert a total of $288$ Ly$\alpha$ emitters into our foreground grism scene. To ensure that all position angles contain
our simulated emission lines, we randomly place our Ly$\alpha$ emitters
within a radius of $1$ arcmin of the field center.

\subsection{Simulation Construction\label{constr}}

We use the aXeSIM \citep{kummel09} software package to simulate 25 independent position angles over a $2048\times2048$ pixel or 14.1 arcmin$^{2}$ field of view. 
aXeSIM was originally designed to simulate grism and direct images for the \textit{Hubble Space Telescope} and a number of modifications were needed to simulate \textit{Roman} data. We modified aXe's instrument parameter file to match \textit{Roman}'s trace and wavelength dispersion. We simulated a wavelength range of $9500-19000$ \AA\ and assumed a constant wavelength dispersion of $265.7$, $10.8$, and $5.6$ \AA\ pix$^{-1}$ for orders 0-0, 1-1, and 2-2, respectively.  We set the zero deviation wavelength of the 1-1 order to be $0.95 \mu$m, which differs from \textit{Roman}'s undeviated wavelength of $\sim1.55 \mu$m.  Given the field of view of the input scene relative to our smaller simulated region, we do not expect this difference to impact our measurement of recovery fractions for simulated Ly$\alpha$ emitters.   

We used the morphological and spectral inputs described in the previous sections to run three separate aXe simulations to form grism images for \textit{Roman}'s 0-0, 1-1, and 2-2 spectral orders. To further facilitate parallelization and to allow for alterations to our Ly$\alpha$ emitter sample, we simulated the grism foreground scene separately from the simulated LAE grism images. 

In Figure \ref{resp}, we show the \textit{Roman} grism response functions that we used in our aXe simulations.  Compared to the science 1-1 order, the 0-0 and 2-2 orders have less-sensitive response functions.  However, these orders can significantly complicate the foreground scene. The 0-0 order is compact with the full $1\mu$m grism range only occupying tens of pixels. A compact 0-0 order from a bright foreground object may be mistaken as an emission line, and we wish to verify that our LAE search does not select these potential contaminants.  The 2-2 order is spread out over $\sim2000$ pixels, about twice the length of the science order.  This means that bright objects well outside of the simulated FOV can have their 2-2 light dispersed into the grism image and contribute to the foreground flux that we need to contend with when searching for high-redshift LAEs.

 In Figure \ref{construct67}, we show an example of our simulated spectral orders and their summation
for a position angle (PA) aligned with the x-axis or PA$=0$ by aXe
convention. Non-zero position angles are not naturally supported by
the aXe simulation software. Thus, we rotated all objects' morphologies
and positions about the field center to simulate position angles not
equal to zero. 

\begin{figure*}[!t]
\centering{}\includegraphics[bb=0bp 2bp 1549bp 513bp,clip,width=18cm]{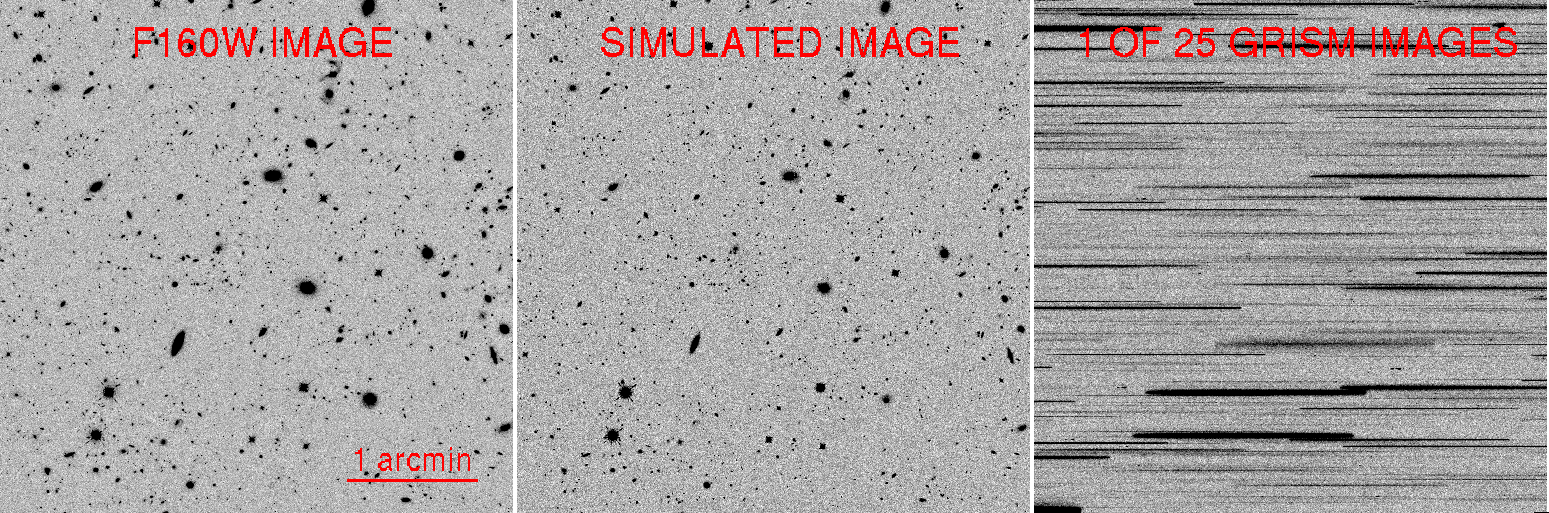}\caption{\textbf{\textit{Left:}} Observed \textit{HST} F160W image. \textbf{\textit{Middle:}}
Our simulated F160W image. \textbf{\textit{Right:}} Our simulated \textit{Roman}
grism image with a position angle aligned with the x-axis. For our
deep field simulations, we consider up to 25 independent position
angles and up to 69.4 hours of total integration time. We use a noise
level of $0.8$ $e^{-}$pix$^{-1}$s$^{-1}$ which is the expected
near-infrared background for the High Latitude Survey.}
\label{grism_fin67}
\end{figure*}

Once all three spectral order simulations were completed for both the foreground and LAE scene, we combined
them and applied a Poisson noise level of $0.8$ $e^{-}$pix$^{-1}$s$^{-1}$
which is the expected near-infrared background for the High Latitude
Survey (see Figure \ref{grism_fin67}). For our deepest simulated \textit{Roman} grism survey, we
explored an observational strategy consisting of $25$ exposures --
each with a unique PA -- observed for $10$ ks giving a total exposure
time of $69.4$ hours. The simulated PAs were selected to uniformly sample 360 degrees, while avoiding 180 degree reflections.  This was done because PA$_{0}$ and PA$_{0}+180$ will have largely the same spectra that are aligned and overlapping, complicating our effort to deblend sources. 

We also investigated two lesser surveys: one
with 25 PAs and a total exposure time of 41.7 hours ($25\times6$
ks) and another with a subset of 15 PAs and the same total exposure time of 41.7
hours ($15\times10$ ks). For brevity, we refer to these three scenarios
as $25\times10$ ks, $25\times6$ ks, $15\times10$ ks.

\subsection{Simulation Caveats}

\textit{Roman}'s wide field slitless spectroscopy with a wavelength
dispersion of $\sim11$ \AA\ per pixel over a $1$ $\mu$m range
is achieved with a compound grism with two diffractive surfaces. This
grism assembly mitigates wavelength-dependent aberrations at the expense
of producing additional unwanted spectral orders. Our simulations
are designed to capture the grism characteristics most relevant for
a deep LAE survey. In particular, we aim to assess \textit{Roman}'s
ability to conduct a deep LAE survey given a realistic foreground
scene. Accordingly, we simulate the spectral off-orders that have
the highest surface brightness (0-0 and 2-2) and ignore out-of-focus
spectral orders (1-0, 1-2, 0-1, 2-1).  Additionally, we do not model
field-dependent distortions of the spectral trace or dispersion which
are expected to have a minor impact on the performance of a deep LAE
survey.

Our simulations are designed to predict the needed on-target time to perform a deep LAE survey, and we do not investigate different dithering/mosaicing strategies. For simplicity, all 25 simulated PAs are assigned the same field center maximizing the area that is covered by all PAs.

\begin{figure}
\begin{centering}
\includegraphics[height=4.75cm]{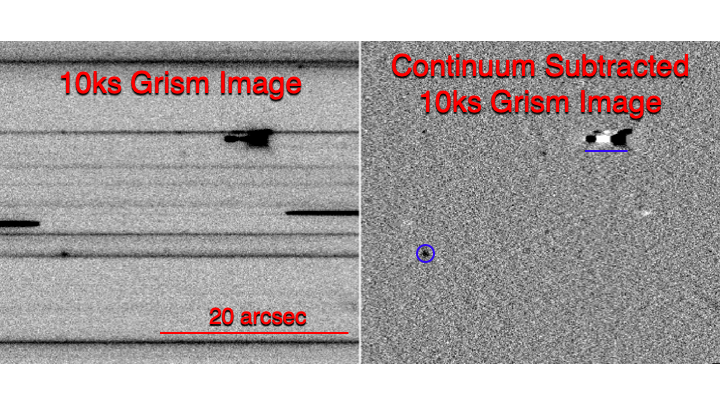}
\vspace{-1.cm}
\caption{We show a small $40\times40''$ region of our simulated $t_{\rm{exp}}=10$ks grism image before and after continuum subtraction.  We have highlighted a foreground emission line galaxy (blue circle) and an artifact caused by the 0-0 order of a bright star (underlined in blue). See Section \ref{subc} for additional discussion of our continuum subtraction method. We demonstrate that our emission line search technique, CUBGRISM, can effectively remove contamination from 0-0 spectra and recover a clean sample of emission line sources.}\label{subcont}
\par\end{centering}
\end{figure}

\section{CUBGRISM \label{cubgrism}}

We use a novel data cube search technique - CUBGRISM - originally
developed for \textit{GALEX} grism data to test our ability to recover
high-redshift \textit{Roman} LAEs. This technique does not require
a broad-band detection and was constructed to produce a line-flux limited sample of emission-line objects. These features are well-tailored
to our goal of characterizing our ability to recover Ly$\alpha$ emitters
given an assumed \textit{Roman} grism deep field observing strategy.

\begin{figure*}[t]
\centering{}\includegraphics[width=18cm]{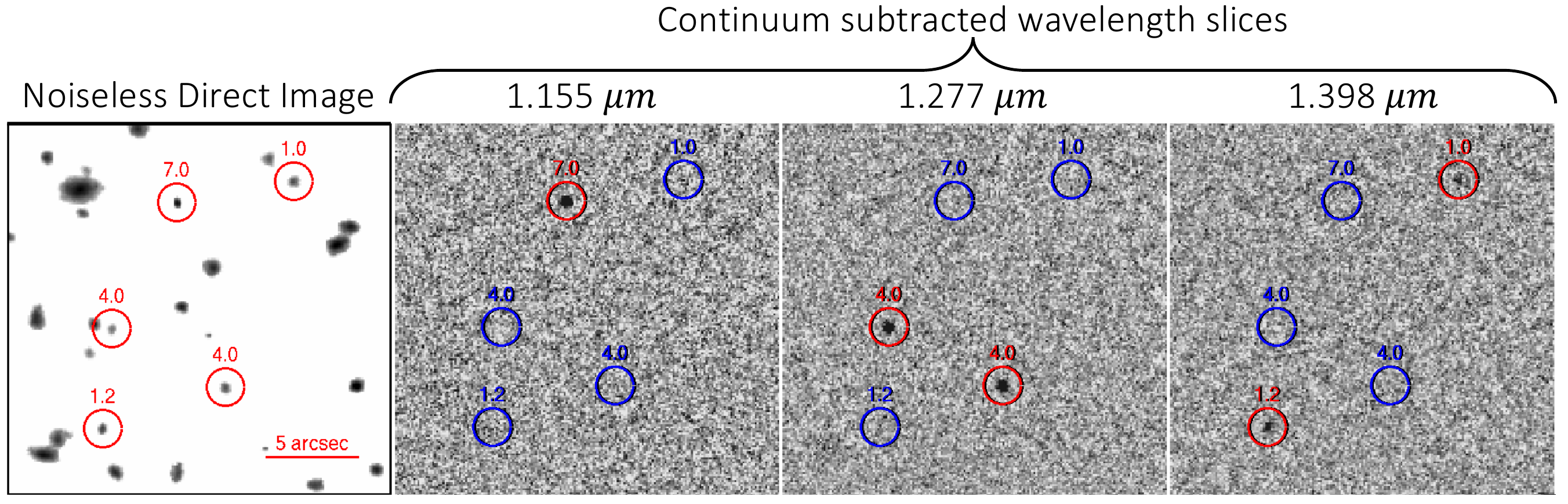}\caption{\label{cube_example}A $20''\times20''$ region of our simulated \textit{Roman}
direct image and three selected wavelength slices from our data cube
constructed from 25 \textit{Roman} grism images. From left to right,
we show the \textit{Roman} direct image, a 1.155 $\mu$m data cube
slice containing 1 simulated $z=8.5$ LAE, a 1.277 $\mu$m data cube
slice containing 2 simulated $z=9.5$ LAEs, and a 1.398 $\mu$m data
cube slice containing 2 simulated $z=10.5$ LAEs. The simulated LAEs
are circled in red. The number associated with each LAE indicates
the line-flux of the source, where the Ly$\alpha$ flux is in units
of $10^{-17}$erg s$^{-1}$cm$^{-2}$. By design continuum sources
are not recovered in our data cube due to our background subtraction
procedure (see Section \ref{cubgrism} for details).}
\end{figure*}

\subsection{Continuum Subtraction\label{subc}}
In preparation for the data cube construction and our emission line search, we subtracted continua from the grism images by applying a $1\times21$ pixel median filter with the $21$ pixel dimension aligned with the spectral trace.  The size of the median filter ($21$ pixels or $231$\AA) is designed to not bias our search for LAEs which will have FHWMs a factor of $\gtrsim20$ smaller. In Figure \ref{subcont}, we demonstrate our continuum subtraction method showing that the remaining sources are primarily emission lines (both simulated LAEs and foreground sources) and 0-0 spectra.

\subsection{Cube Construction \label{cconstr}}

In \citet{barger12,wold14,wold17}, we describe CUBGRISM (previously
referred to as the data cube search) in detail and demonstrate our
method for converting multiple slitless spectroscopic images into
a three-dimensional data cube with two spatial axes and one wavelength
axis. Here, we provide a brief overview of this process with emphasis
on alterations needed for \textit{Roman} grism images.

For each simulated grism exposure, we know the spectral trace and
wavelength dispersion, and this allows us to extract a spectrum for
each spatial pixel position, thus forming an initial data cube. A
data cube constructed from a single slitless spectroscopic image may
suffer from overlapping spectra caused by neighboring objects that
are oriented in line with the dispersion direction. For our simulated
\textit{Roman} deep field, we have multiple position angles - one
per exposure - and objects that overlap in one position angle are
unlikely to overlap in another position angle. This allows us to disentangle
overlaps by constructing data cubes for each exposure and then 
combining these initial data cubes, applying a 5$\sigma$ cut to remove
contamination from overlapping sources. In more detail, we take our library of data cubes ($N=25$ for our deep survey) and for each spaxel location, we measure the variation in flux values, quantified by the normalized median absolute deviation.  Any spaxel that deviates from the median by more than $5\sigma$ is eliminated and the remaining spaxels are used to compute the median flux value which is then used to populate the final data cube. 

This procedure results in a final data cube
that has a wavelength step of 10 \AA\ and wavelength coverage of approximately 1.0--1.5 $\mu$m.  We designed the wavelength slices to have a wavelength extent
that matches the spectral resolution of the \textit{Roman Space Telescope}.
We are interested in our ability to recover high-redshift Ly$\alpha$
emitters ($z=7.5$--$10.5$), and accordingly, we 
artificially limit the wavelength range of the output data.

\subsection{Line Search \label{csub}}

The final  \textit{Roman} data cubes cover a wavelength range
of 1.00 to 1.49 $\mu$m or a Ly$\alpha$ redshift range of $z=7.24-11.25$.
For each wavelength slice, we used SExtractor \citep[SE;][]{bertin96} to
identify $>7.5\sigma$ detections using SNR-optimised aperture photometry.  We computed an optimal aperture size of $D=0.4''$ by determining the aperture that maximizes the median Ly$\alpha$ data-cube flux relative to the measured aperture noise \citep[for details on SNR-optimal apertures, see][]{gawiser06}.  As discussed in the next section, this level of significance ($>7.5\sigma$) is needed to minimize false detections within the cube.

\subsection{False detections\label{false}}

CUBGRISM enables a blind search for Ly$\alpha$ emitters -- with
no continuum detection required -- at the expense of requiring a
PA-rich observational strategy to accurately isolate emitter locations.
We favor this continuum-independent reduction strategy because our
science goals require us to compare \textit{Roman }Ly$\alpha$ populations
at cosmic dawn to blind Ly$\alpha$ narrow-band searches just outside
the reionization epoch at $z\lesssim7$ \citep[][]{ouchi08,ouchi10,shibuya12,konno14,konno18,matthee15,santos16,ota17,itoh18,hu19,hu21,tilvi20,wold22}.

However, for reduced PA observational strategies, a potential drawback
of our technique is the inability to isolate source locations resulting
in spurious detections.

CUBGRISM has been used successfully on \textit{GALEX} grism deep fields
observed with hundreds of PAs to identify populations of low-redshift
Ly$\alpha$ emitters. These far-ultraviolet and near-ultraviolet surveys
found a high optical-spectroscopic confirmation rate of $85\%$ \citep{wold14,wold17}.
In addition to a $4\sigma$ detection within data cube, these \textit{GALEX}
studies also visually inspected profile-weighted one-dimensional spectra
to help eliminate spurious detections. For our Roman simulations,
we did not implement a hard-to-quantify visual inspection step. Instead,
we verify that our adopted CUBGRISM extraction parameters result in
a manageable contamination rate ($\lesssim15\%$) which would likely
be improved with a visual inspection or machine learning step.

\begin{deluxetable*}{ccccc} 
\tablecolumns{5} 
\tablewidth{0pc} 
\tablecaption{\textit{Roman} Grism Survey Performance}
\tablehead{ 
\colhead{Total} \vspace{-0.2cm} & \colhead{Number} & \colhead{Average survey} & \colhead{False} & \colhead{Figure \ref{survey_complete}}\\
\colhead{Exposure} \vspace{-0.2cm} & \colhead{of} & \colhead{Flux Limit\tablenotemark{\rm{\footnotesize{a}}}} & \colhead{Detection} & \colhead{symbol}\\
\colhead{Time (hrs)} & \colhead{PAs} & \colhead{($\times10^{-18}$ erg s$^{-1}$cm$^{-2}$)} & \colhead{Fraction\tablenotemark{\rm{\footnotesize{b}}}} & \colhead{color}}
\startdata 
69.4 & 25 & 8.2$\pm$0.7 & 0.137$\pm$0.048 & red\\
41.7 & 25 & 10.8$\pm$0.6 & 0.074$\pm$0.069 & green\\
41.7 & 15 & 10.4$\pm$0.5 & 0.058$\pm$0.055 & blue\\
41.7 & 10 & 10.2$\pm$0.6 & 0.678$\pm$0.035 & \nodata
\enddata  
\tablenotetext{\rm{\footnotesize{a}}}{Survey flux limits are computed from the $50\%$ completeness threshold for all simulated LAEs $z=7.5$--$10.5$.}
\tablenotetext{\rm{\footnotesize{b}}}{The false detection fraction is defined as $F/(F+E)$ where $F$ is the number of recovered spurious objects within the $z_{\rm{Ly\alpha}}=7.25$--$8.75$ redshift range and $E$ is the number of expected LAEs over the same redshift range assuming no number density evolution from the $z\sim7$ LAGER survey.}
 \vspace{-0.7cm}
\label{table1}
\end{deluxetable*}

\begin{figure}
\centering{}\includegraphics[bb=79.2bp 69.3204bp 712.8bp 524.854bp,clip,height=6cm]{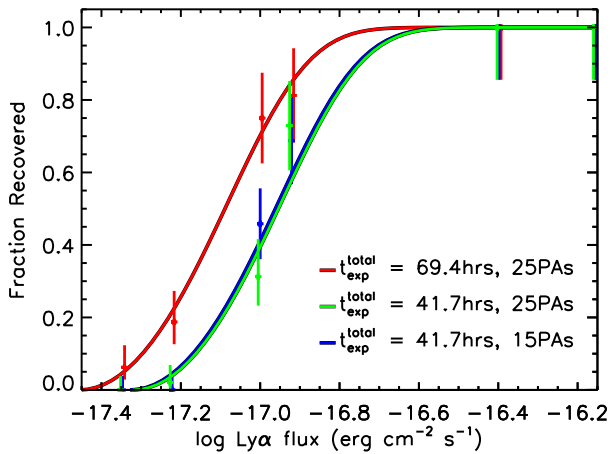}\caption{\label{survey_complete} The fraction of recovered Ly$\alpha$ emitters
as a function of line-flux. Error-bars show the $1\sigma$ Poisson
error. The red, green, and blue curves show examples of how our survey
completeness varies with different observational strategies assuming
a $0.8$ $e^{-}$pix$^{-1}$s$^{-1}$ background noise level typical
of the High Latitude Survey (see Section \ref{results} for details).}
\end{figure}

For a deep Roman grism survey, we aim to detect the Ly$\alpha$ number
density decline caused by the increasing opacity of the IGM and the
onset of reionization. As a baseline for determining the rate of false detections, we define the false detection fraction as $F/(F+E)$ where $F$ is the number of recovered spurious objects within the $z_{\rm{Ly\alpha}}=7.25$--$8.75$ redshift range and $E$ is the number of expected LAEs over the same redshift range assuming no number density evolution from the $z\sim7$ LAGER survey \citep[e.g.,][]{wold22,hu19}. 
 Given our survey flux limits, we expect a
$69.4$ hr survey to detect $N\sim400$ LAEs while our $41.7$ hr surveys should detect $N\sim150$ LAEs over a nominal deep grism survey area of $1$ deg$^{2}$ (see Section \ref{results} for details).

It is not possible to simulate a larger $\sim1$ deg$^{2}$ FOV because of the limited
\textit{HST} near-infrared imaging data and our COSMOS observation-based
simulations. Instead, we assume that our limited FOV is representative
and perturb our simulations with different realizations of Poisson
noise. For each realization, we find the number of false detections
over a wavelength range of 1.00 to 1.19 $\mu$m, which corresponds
to the wavelength range of observed $z=7.25$--$8.75$ Ly$\alpha$ emission.
In total, we look for false detections over an effective survey area
350 arcmin$^{2}$ and consider all three survey scenarios: $25\times10$
ks, $25\times6$ ks, $15\times10$ ks.

For all three survey configurations, we find that a $7.5\sigma$ detection threshold
can keep the false detection rate at $\lesssim15\%$.  In Table \ref{table1}, we summarize these results with their $1\sigma$ Poisson errors.  We caution that using $N\ll15$ PAs will result in reduced performance of our CUBGRISM reduction strategy.  Using the same procedure as employed with the $25\times10$
ks, $25\times6$ ks, and $15\times10$ ks surveys, we estimate that a $10\times15$ ks PA survey would result in a $\sim70\%$ false detection rate. Reduced performance with a PA-poor survey is expected because of CUBGRISM's method of disentangling overlapping spectra (described in Section \ref{cconstr}) which requires many PAs to measure the uncontaminated flux at each $x, y, \lambda$ position.

We note that our false detection results
do not account for foreground emission-line contamination. We assume
that a deep \textit{Roman} grism survey will be located in a field
that contains deep ancillary imaging data that are capable of efficiently
identifying foreground objects. As performed in the \textit{GALEX}
grism and narrow-band Ly$\alpha$ surveys, the false detection rate can be further investigated by independently confirming a representative subset
of identified \textit{Roman} grism Ly$\alpha$ emitters.

\section{Results \label{results}}
Given our generated realistic foreground scene and our reduction strategy, we now determine the recovery
fraction of simulated Ly$\alpha$ emitters as a function of line-flux,
redshift, and number of position angles. In Figure \ref{cube_example},
we demonstrate this Ly$\alpha$ search by highlighting a small $20''\times20''$
region of our deepest simulation that contains 5 LAEs with different
line fluxes and redshifts. For each wavelength slice, we search for LAEs and record the Ly$\alpha$ recovery fraction as a function of line flux.

As discussed in Section \ref{constr}, we focused our
investigation on three different deep-field scenarios.

\begin{enumerate}
\item 25 PAs with a total exposure time of 69.4 hours ($25\times10$ ks)
\item 25 PAs with a total exposure time of 41.7 hours ($25\times6$ ks)
\item 15 PAs with a total exposure time of 41.7 hours ($15\times10$ ks)
\end{enumerate}
For each scenario, we show the fraction of recovered LAEs as a function
of line-flux in Figure \ref{survey_complete}. Following \citet{negrello13},
we fit our recovery fractions with an error function
\begin{equation}
R(f_{{\rm {Ly}\alpha}})={\rm {erf}}[(logf_{{\rm {Ly}\alpha}}-logf)^{2}/\gamma^{2}]
\end{equation}
where $logf$ and $\gamma$ are free parameters that control the location
of the zero-to-one transition and the rate of this transition, respectively.
We find that the $\gamma$ parameter is very similar for all curves,
and we fix its value to the $25\times10$ ks best-fit value of $\gamma=0.52$.
Based on these best-fit functions, we find that our three
scenarios are $50\%$ complete at a Ly$\alpha$ line fluxes of $8.2$,
$10.8$, and $10.4\times10^{-18}$ erg s$^{-1}$cm$^{-2}$,
respectively. This is consistent with the survey depth increasing
proportionally to the square root of the total exposure time. In Table \ref{table1}, we list our surveys' flux completeness with $1\sigma$ errors computed with a Monte Carlo (MC) procedure that perturbs our completeness data-points by Poisson random deviates.  This MC procedure is used to create $N=10,000$ perturbed completeness curves and the measured variation in the survey completeness is used to estimate the $1\sigma$ error.

We investigate \textit{Roman}'s Ly$\alpha$
survey depth as a function of redshift for our deepest $25\times10$
ks survey configuration. Binning by line flux and redshift reduce the number of simulated LAEs per bin by a factor of four when compared to binning by line flux alone. To reduce the Poisson error in our recovery fraction results, we have produced four Ly$\alpha$ samples while keeping our foreground scene the same.  For each Ly$\alpha$ sample, we independently select Ly$\alpha$ parameters as described in Section \ref{lya}.  This includes randomly selecting their positions.  For each run, the Ly$\alpha$ sample is added to the foreground scene and Poisson noise is added. We combine Ly$\alpha$ recovery results from all four runs to compute the final completeness curves.

\begin{deluxetable*}{cccc} 
\tablecolumns{4} 
\tablewidth{0pc} 
\tablecaption{Deep \textit{Roman} Grism Survey Redshift Performance}
\tablehead{
\colhead{Redshift} \vspace{-0.2cm} & \colhead{Flux Limit\tablenotemark{\rm{\footnotesize{a}}}} & \colhead{Luminosity Limit\tablenotemark{\rm{\footnotesize{a}}}} & \colhead{Figure \ref{zzcomplete_z}}\\
\colhead{} & \colhead{($\times10^{-18}$ erg s$^{-1}$cm$^{-2}$)} & \colhead{($\times10^{42}$ erg s$^{-1}$)} & \colhead{symbol color}}
\startdata 
7.5 & 13.6$\pm$0.8 & 9.1$\pm$0.6 & magenta\\
8.5 & 8.3$\pm$0.4 & 7.4$\pm$0.4 & cyan\\
9.5 & 7.5$\pm$0.4 & 8.6$\pm$0.4 & yellow\\
10.5 & 6.9$\pm$0.4 & 9.9$\pm$0.6 & orange
\enddata  
\tablenotetext{\rm{\footnotesize{a}}}{Survey flux limits are computed from the $50\%$ completeness threshold.}

\label{table2}
\end{deluxetable*}

\begin{figure*}
\vspace{-1.0cm}
\centering{}\includegraphics[bb=99bp 69.3204bp 712.8bp 524.854bp,clip,height=6cm]{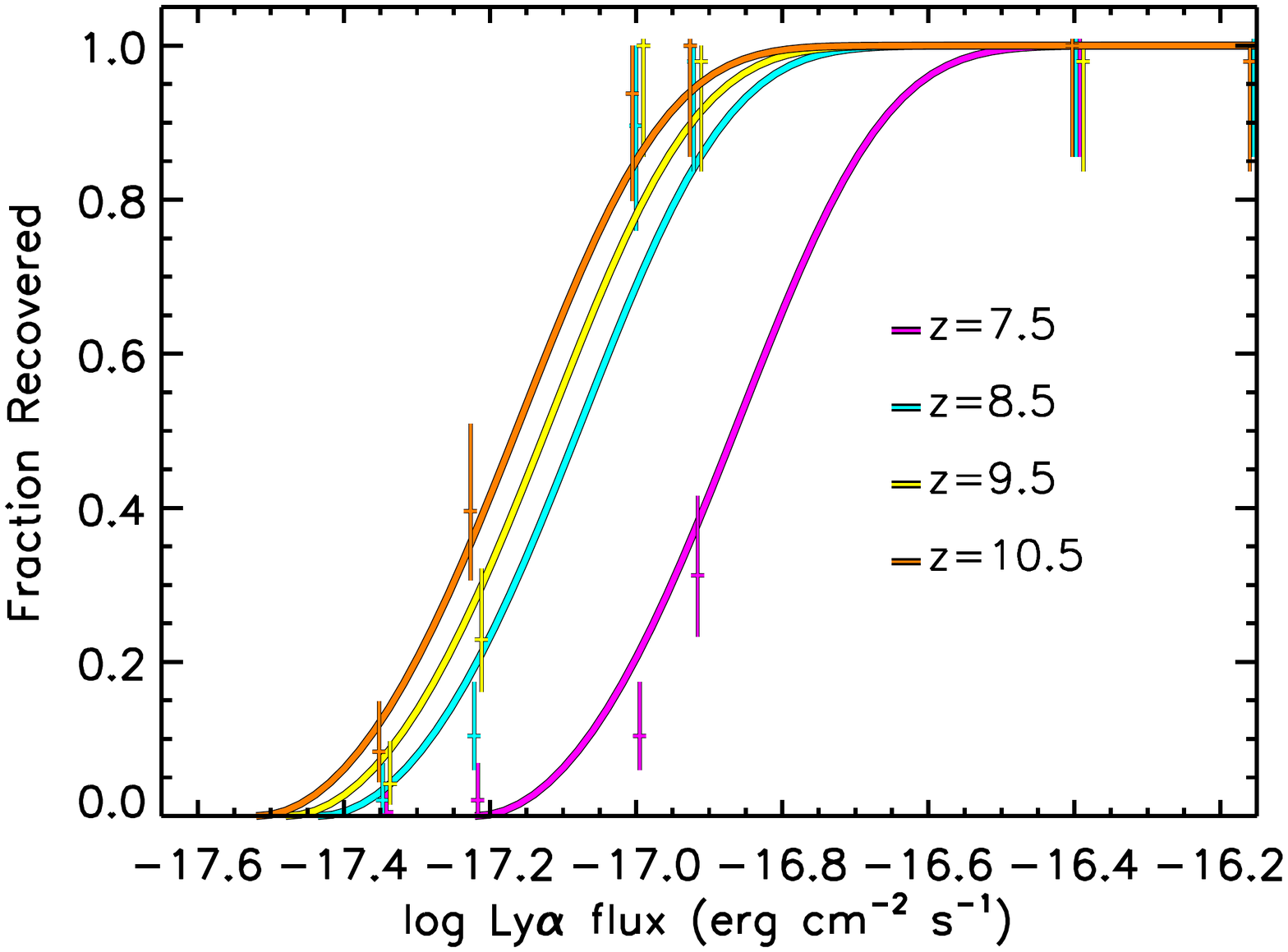}\includegraphics[bb=79.2bp 69.3204bp 712.8bp 524.854bp,clip,height=6cm]{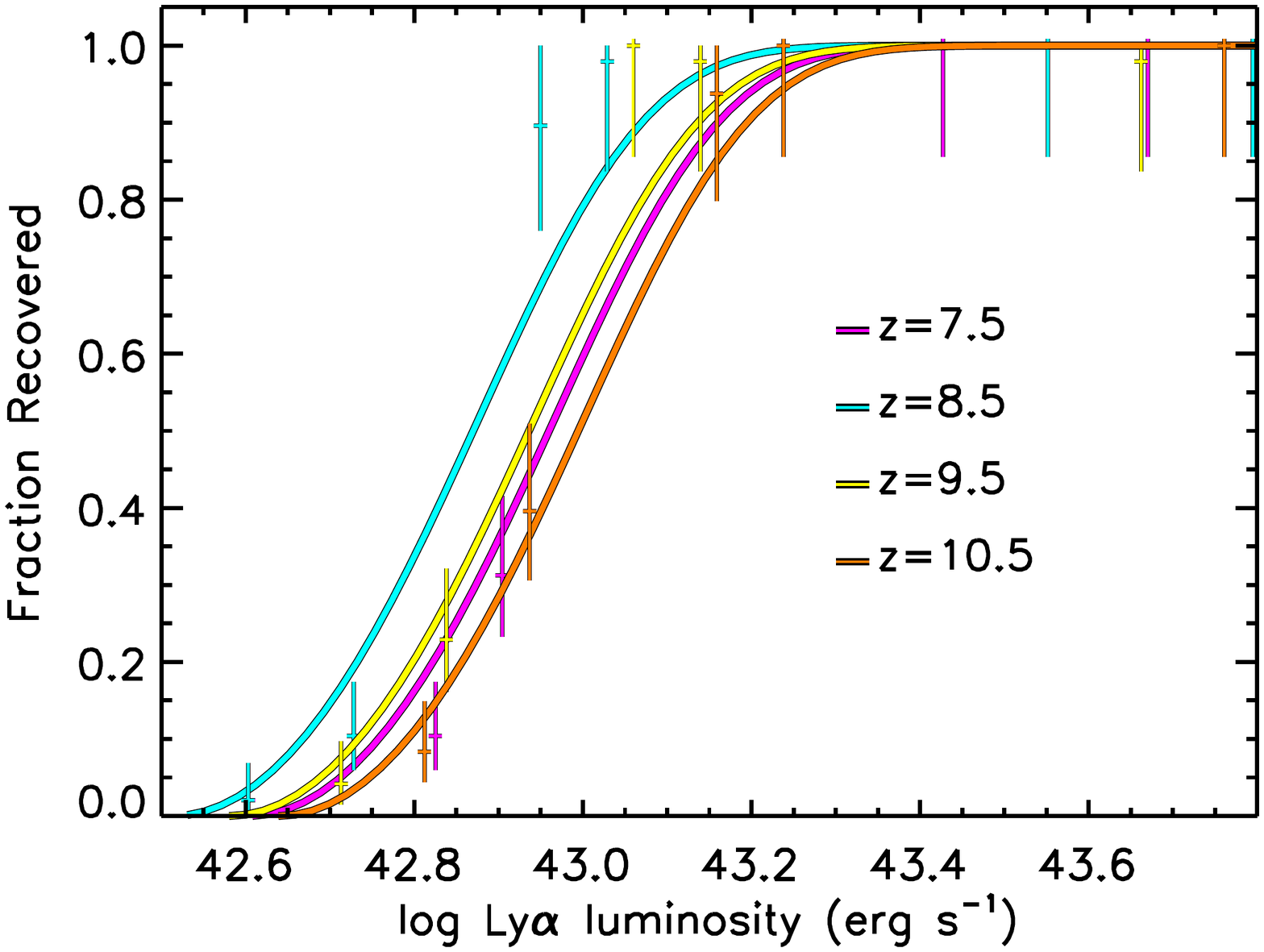}\caption{\label{zzcomplete_z}\textbf{\textit{Left:}} For our deepest grism
simulation ($t_{{\rm {exp}}}=69.4$ hrs), we show Ly$\alpha$ flux
completeness curves for different emitter redshifts. \textbf{\textit{Right:}}
Same as the left figure except completeness is plotted versus Ly$\alpha$
luminosity. Across all redshift bins, we find that our survey can recover $L_{{\rm {Ly}}\alpha}>10^{43}$ erg s$^{-1}$ LAEs at the $50\%$ completeness level (see Table \ref{table2}).   }
\end{figure*}

In Figure \ref{zzcomplete_z}, we show these redshift-dependent Ly$\alpha$ recovery fractions. From $z=7.5$ to $10.5$, we find that our
survey is $50\%$ complete at Ly$\alpha$ line fluxes of $13.6$, $8.3$, $7.5$, and $6.9\times10^{-18}$
erg s$^{-1}$cm$^{-2}$, respectively. The decline in survey depth
at $z=7.5$ is due to the grism's response function (see Figure \ref{resp}).
From a wavelength of 1.16 to 1.03 $\mu$m or a redshift range of $z_{{\rm {Ly}\alpha}}=8.5$
to $7.5$, the grism response declines by a factor of $1.6$.

In the right panel of Figure \ref{zzcomplete_z}, we show that the
reduced cosmological dimming at $z=7.5$ makes up for the
decline in the grism response function allowing for the recovery of
$L_{{\rm {Ly}}\alpha}>10^{43}$ erg s$^{-1}$ LAEs across all redshift bins. In terms of Ly$\alpha$
luminosity, our simulated \textit{Roman} survey is most sensitive
at $z=8.5$ with a $50\%$  survey completeness at $7.4\times10^{42}$ erg s$^{-1}$.   We summarize the $25\times10$ ks survey's redshift performance in Table \ref{table2}.

Considering a nominal deep survey area of $1$ deg$^{2}$ with a $z=7.25$--$8.75$ redshift range and assuming no number density evolution from the $z\sim7$ LAGER survey \citep{wold22}, we estimate a survey sample size of 395 LAEs. To compute this quantity, we consider three redshift slices: $z=7.5$, $8.0$, and $8.5$ each with a $z\pm0.25$ extent.  For each slice, we compute the Ly$\alpha$ number density by integrating the $z\sim7$ LAGER LF down to our redshift dependent $50\%$ completeness limits, where the $z=8.0$ completeness is interpolated from our $z=7.5$ and $8.5$ measurements. Given the slice volumes (where $V_{\rm{Total}}=1.1\times10^7$ Mpc$^3$), the total number of LAEs is found to be 395 within an area of $1$ deg$^{2}$ and a $z=7.25$--$8.75$ redshift range.  Any significant decline in this sample size is a signature of the increasing opacity of the IGM within the reionization epoch.  To give an estimate of this decline, we note that from a redshift of $z=7$ to $8$ the number density of star-forming galaxies declines by a factor of $\sim2$ \citep[as traced by the evolution of the UV LFs][]{bouwens15}, indicating that a $\gtrsim2$ Ly$\alpha$ number density decline over the same redshift range may be attributed to increasing IGM opacity during the reionization epoch.  

We note that there are examples of spectroscopically confirmed $z\gtrsim8$ LAEs within our detection limits \citep[e.g.,][]{oesch15,zitrin15,larson18,tilvi20,jung20},
indicating that these sources exist and can be recovered with a large-volume survey. With JWST observations the number of spectroscopically confirmed $z\gtrsim8$ sources are increasing rapidly with the highest redshift LAE currently at a redshift of $z=10.6$ \citep{bunker23}.  However, these existing $z\gtrsim8$ LAE samples have a relatively small survey volume or are pre-selected based on photometric redshifts. 

For example, \citet{tilvi20} found one L$_{\rm{Ly}\alpha}>10^{43}$ erg s$^{-1}$ LAE at $z=7.7$  over a $\sim 4.5\times10^{4}$ Mpc$^3$ narrow-band survey. Naively scaling the implied $z=7.7$ LAE number density by the volume of our nominal deep \textit{Roman} grism survey, implies 1) a Roman sample size of $230$ LAEs, 2) a high IGM Ly$\alpha$ transmission, and 3) a small neutral hydrogen fraction. Clearly, redshift evolution, cosmic variance, and Poisson fluctuations make this estimate highly uncertain.  Considering just the $1\sigma$ Poisson uncertainty \citep{gehrels86}, allows for \textit{Roman} sample sizes ranging from $40$ to $770$ LAEs assuming a $1$ deg$^{2}$ FOV and a $z=7.25$--$8.75$ redshift range. To overcome these uncertainties, we need a deep \textit{Roman} survey to conduct a blind search for strong Ly$\alpha$ emitters over a large volume.  This will allow us to make a direct comparison to the number density evolution prior ($z\sim5.7$ via ground-based NB Ly$\alpha$ surveys) and during the reionization epoch ($z\gtrsim7$ via a deep \textit{Roman} grism survey).   

At a redshift of $z=8$, models of reionization predict significantly different volume-averaged neutral hydrogen fractions, $x_{HI}$.  For example, at this redshift \citet{naidu20} predict an almost fully neutral universe with $x_{HI}=87\%$, while the \citet{finkelstein19} model predicts a much lower neutral fraction of $x_{HI}=35\%$. One of the main differences in these models is whether massive galaxies or more numerous low-mass galaxies account for the majority of ionizing flux. With a deep \textit{Roman} grism survey, we will obtain observational constraints on the evolution of the $z\sim8$ Ly$\alpha$ LF and be able to distinguish between a rapid neutral-to-ionized transition dominated by relatively massive galaxies  or a gradual transition dominated by more numerous low-mass galaxies.

A deep \textit{Roman} survey will also allow us to study the shape of the Ly$\alpha$ luminosity function.  Our completeness simulations indicate that a deep \textit{Roman} grism survey can achieve
Ly$\alpha$ line depths comparable to the deepest $z\gtrsim7$ NB
surveys \citep[e.g.,][]{clement12,hibon12,shibuya12,konno14,itoh18,ota17,hu19,tilvi20,wold22}.
Based on these $z\sim7$ Ly$\alpha$ surveys, this depth will allow
us to measure the shape of the Ly$\alpha$ luminosity function at the reionization epoch. Previous studies have suggested the bright-end of the luminosity function may evolve less rapidly due to bright LAEs preferentially residing within large ionized bubbles \citep[][]{matthee15,hu21,taylor20,taylor21,ning22} and a deep \textit{Roman} grism survey will allow us to study the bright-end of the LF at $z\gtrsim8$ for the first time.

\section{Summary}

We investigated \textit{Roman}'s potential to conduct a blind search
for Ly$\alpha$ galaxies at cosmic dawn using a multi-position-angle
observational strategy. We produced a realistic \textsl{Roman} WFI
grism foreground scene based on observational constraints from the
COSMOS field. We also simulated Ly$\alpha$ galaxies spanning our
redshift and line-flux range of interest. We showed how a novel data
cube search technique -- CUBGRISM -- originally developed for \textit{GALEX}
grism data can be applied to \textit{Roman} grism data to produce
a Ly$\alpha$ flux-limited sample without the need for a continuum
detection. Given our adopted reduction technique, we investigated
the impact of altering the number of independent position angles and
total exposure time. Our results indicate that a proposed deep \textit{Roman}
grism survey can achieve Ly$\alpha$ line depths comparable to the
deepest $z=7$ NB surveys, allowing us to study the evolution of Ly$\alpha$
populations and infer the ionization state of the intergalactic medium
at cosmic dawn. 

\noindent \acknowledgments{IGBW is supported by an appointment to the NASA Postdoctoral Program at the Goddard Space Flight Center. The material is based upon work supported by NASA under award number 80GSFC21M0002, and via the WFIRST Science Investigation Team contract NNG16PJ33C, `Studying Cosmic Dawn with WFIRST'.

This work is based on observations taken by the 3D-HST Treasury Program (GO 12177 and 12328) with the NASA/ESA HST, which is operated by the Association of Universities for Research in Astronomy, Inc., under NASA contract NAS5-26555.}

\bibliographystyle{aasjournal} 
\bibliography{wold}

\end{document}